\documentclass[epj]{webofc}
\usepackage[varg]{txfonts}   % Web of Conferences font
\wocname{EPJ Web of Conferences}
\woctitle{ICNFP 2015}
\newcommand{\eq}[1]{\begin{align} #1 \end{align}}
\begin{document}
\selectlanguage{english}
\title{High temperature Bose-Einstein condensation}
%
% subtitle is optional

\author{Viktor Begun\inst{1}\fnsep\thanks{\email{viktor.begun@gmail.com}}
%         \and
%        Second author\inst{2} \and
%        Third author\inst{3}
        % etc.
}

 \institute{Institute of Physics, Jan Kochanowski University, Kielce, Poland
%\and
%           The second here
%\and
%           Last address
}

\abstract{The indications of a possible pion condensation at the
LHC are summarized. The condensation is predicted by the
non-equilibrium hadronization model for 2.76~TeV Pb+Pb collisions
at the LHC. The model solves the proton/pion puzzle and reproduces
the low $p_T$ enhancement of the pion spectra, as well as the
spectra of protons and antiprotons, charged kaons, $K^0_S$,
$K^*(892)^0$ and $\phi(1020)$. The obtained parameters allow to
estimate the amount of pion condensate on the level of 5\% from
the total number of pions at the LHC. The condensate is located at
$p_T<250$~MeV.}
\maketitle
 \section{Introduction} \label{intro}
Bose-Einstein condensation (BEC) was predicted right after the
introduction of Bose statistics~\cite{BEC}. However, it took 70
years to observe BEC experimentally~\cite{BEC-1}. The main
difficulty for obtaining BEC in a gas of atoms is that they form a
liquid or a solid before reaching BEC. It can be avoided, if
extremely low densities and temperatures are achieved. This was
done only after the development of the corresponding technologies,
and the leaders of the two groups that created the condensate
received the Nobel Prize in 2001~\cite{BEC-2}. It might be the
reason for the widespread delusion that BEC means low temperatures
and atomic gases only.
However, all conventional two-quark mesons are bosons and may
condense. The temperatures that are reached in high-energy
collision are of the MeV scale. It corresponds to $\sim 10^{10}$
Kelvin and is far above the BEC temperatures for cold atoms.
The ratio of the BEC temperature in atomic gases, $T_C(A)$, to
that in the pion gas, $T_C(\pi)$, is even larger, because one
should take into account a possible size of the system and the
masses of the particles. One obtains $10^{12}$ times higher
temperature for the gas of pions~\cite{Begun:2008hq}:
 \eq{
  \frac{T_C(\pi)}{T_C(A)}
  ~\simeq~ \frac{m_A}{m_{\pi}}\left(\frac{r_A}{r_{\pi}}\right)^2 ~\simeq~ \frac{m_A}{m_{\pi}}10^{10}
  ~\simeq~ 10^{12}~.
 }
Besides of much higher temperature, the properties of a BEC of
mesons could be very different from the low-temperature BEC of
atoms. The most obvious differences are the much smaller volumes,
much higher densities, and different interaction forces involved
in the formation of the high temperature BEC. The clear advantage
of the pion BEC is that pion system is in the form of gas at
freeze-out.
The reason to consider pion BEC now is the recent data from Pb+Pb
collisions at $\sqrt{s_{NN}}=2.76$~TeV energies at Large Hadron
Collider (LHC), that contain the set of puzzles.
 \begin{itemize}
 \item The predictions of the hadron-resonance gas (HRG)
 were too high for ratios to pions, especially for proton to pion ratio~\cite{Abelev:2013vea}.
 \item
 The best fit of the LHC data by the HRG still gives nearly three standard deviations for
 protons~\cite{Stachel:2013zma}.
 \item
 The low-transverse-momentum pion spectra show up to 50\% enhancement compared to hydrodynamic
 models~\cite{Abelev:2013vea}.
 \item
 The temperature obtained in HRG at the LHC falls out from the freeze-out line deeply in the
 hadronic phase~\cite{Cleymans:2014xha}.
 \end{itemize}

The HRG worked well for smaller collisions energies. Pions and
protons are among the most abundant particles. Therefore the
problems with their description look suspicious.
There are several solution proposed to explain the proton to pion
ratio. The hadronization and freeze-out in a chemical
non-equilibrium \cite{Petran:2013lja}. The separate freeze-out for
strange particles~\cite{Chatterjee:2014lfa}. An incomplete list of
hadrons~\cite{Noronha-Hostler:2014aia}, and hadronic rescattering
in the final stage~\cite{Becattini:2012xb}. However, none of them
is commonly accepted yet.

There are deep physical reasons for the non-equilibrium and pion
condensation at the LHC. It can be due to fast expansion and
overcooling of the QGP~\cite{Csorgo:1994dd,Shuryak:2014zxa}, or
due to gluon condensation in Color Glass
Condensate~\cite{Blaizot:2011xf}, and subsequent hadronization of
the low $p_T$ gluons into low $p_T$ pions~\cite{Gelis:2014tda}.
The non-equilibrium hadronization can explain the measured
particle ratios~\cite{Petran:2013lja}, and also the
spectra~\cite{Begun:2013nga,Begun:2014rsa} very well. The obtained
parameters indicate the possibility that $5\%$ of the total number
of pions can be in BEC at the LHC~\cite{Begun:2015ifa}.
The calculations of the pion spectra with the assumption of BEC
show that  the maximal transverse momentum that the BEC can obtain
due to the expansion of the fireball is
$p_T<250$~MeV~\cite{Begun:2015ifa}. It is the same region where
ALICE Collaboration observes the emission of 23\% of pions from a
coherent source~\cite{Abelev:2013pqa}.

\section{The non-equilibrium model}\label{sec-1}
The thermodynamic motivation behind the non-equilibrium HRG (NEQ)
is the following. The fast expansion of the fireball may cause
it's overcooling, so that the number of quarks and anti-quarks is
larger than their equilibrium value at the reached temperature.
Therefore, the subsequent freeze-out may lead to the formation of
mesons and baryons with the multiplicities that are higher than
the corresponding equilibrium numbers at the given temperature.
The phase-space distribution of the primordial particles in NEQ is
similar to the usual HRG:
\eq{
 f_i ~=~ g_i \int \frac{d^3p}{(2\pi)^3}
 \frac{1}{\gamma_i^{-1}\exp(\sqrt{p^2+m^2}/T)\pm1},
 }
where $g_i$ is the degeneracy factor, $p-$momentum, $m-$mass,
$T-$temperature, and
 \eq{\label{f-i}
 \gamma_i ~=~
 \gamma_q^{N^i_q+N^i_{\bar q}}
 \gamma_s^{N^i_s+N^i_{\bar s}}  \exp \left( \frac{ \mu_B B_i  + \mu_Q Q_i  + \mu_S
 S_i}{T}\right)~.
}
The $\gamma_i$ contains the chemical potentials and corresponding
conserved charges ($Q$, $B$, $S$) of a particle $i$ in equilibrium
HRG (EQ), and the non-equilibrium parameters $\gamma_q$,
$\gamma_s$. The $N^i_q$, $N^i_{\bar{q}}$ and $N^i_s$,
$N^i_{\bar{s}}$ are the numbers of light $(u,d)$ and strange $(s)$
quarks and anti-quarks in the $i-$th hadron, correspondingly. Note
the plus sign between the number of quarks and anti-quarks, in
contrast to the usual conserved charges, that are defined as the
difference between particles and anti-particles.
One can exponentiate $\gamma_q$ and $\gamma_s$ and see that each
particle in NEQ obtains it's own chemical potential, which can be
also recalculated into a different temperature. Therefore NEQ is a
particular case of the multiple-freeze out model, where each
particle has it's own freeze-out, but the relation between them is
fixed by Eq.~(\ref{f-i}).
The NEQ is advocated by J.~Rafelski and collaborators since a long
time, see Ref.~\cite{Rafelski:2015cxa} for the review. It is
implemented in the SHARE model~\cite{Torrieri:2004zz} and allows
for the HRG fit of the particle multiplicities and ratios. The
THERMINATOR Monte-Carlo event generator~\cite{Chojnacki:2011hb}
inherits the same particle and decay list and allows for further
calculation of the particle spectra from the Cooper-Frye formula
at the freeze-out hypersurface $\Sigma_\mu$:
\eq{
 \frac{dN}{dy d^2p_T} = \int d \Sigma_\mu p^\mu f(p \cdot u), \qquad\quad
 t^2 = \tau_f^2 + x^2 + y^2 + z^2, \qquad\quad x^2 + y^2 \leq r^2_{\rm
 max}~,
}
where $\tau_f$ and $r_{\rm max}$ are the characteristic time and
radius of the hypersurface.
There are several hypersurfaces implemented in THERMINATOR
following the paper~\cite{Broniowski:2001we}. For the LHC we use
Cracow model, that assumes the Hubble-like flow of particles:
$u^\mu = x^\mu/\tau_f$, because it fits the data the best.
At the LHC $\mu_B\simeq\mu_Q\simeq\mu_S\simeq 0$ due to large
collision energy compared to the rest mass of the colliding
nuclei. Therefore, EQ model has two parameters left - temperature
and volume, while NEQ has also $\gamma_q$ and $\gamma_s$. The
measured set of particles, that was used
in~\cite{Begun:2013nga,Begun:2014rsa,Begun:2014aha,Begun:2015ifa},
includes 6 to 8 multiplicities depending on centrality. It allows
to determine both the parameters in EQ and in NEQ.

\begin{figure}[h]
 \centering
 \includegraphics[width=0.48\textwidth]{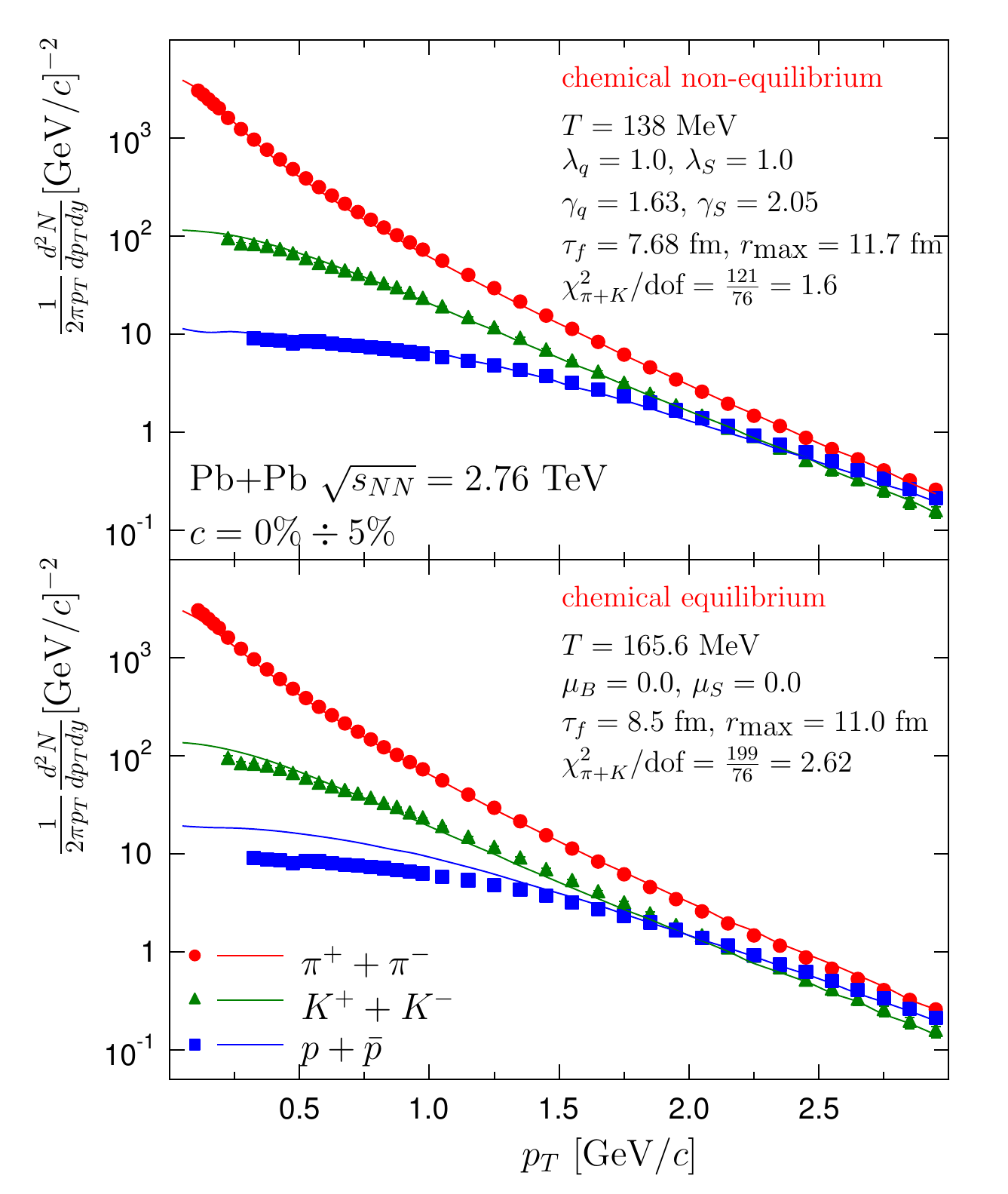}%fig_c0005}
 \hspace{0.2cm}
 \includegraphics[width=0.48\textwidth]{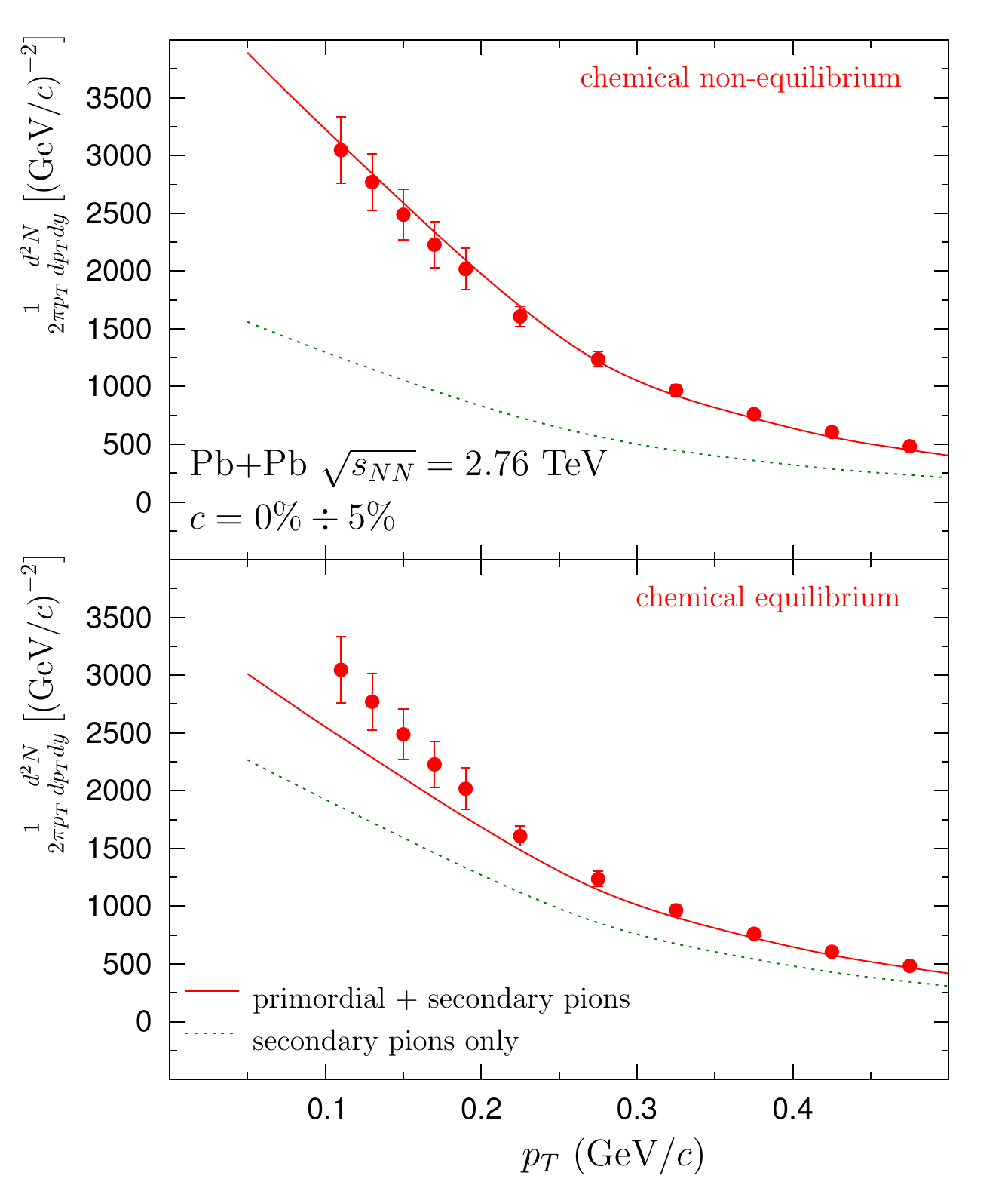}%fig_c0005_l}
 \caption{The spectra of pions, kaons and protons (left) in NEQ (up) and EQ (down), and the same only for pions in linear scale (right).
 The figures are from Ref.~\cite{Begun:2013nga}.}
 \label{fig-1}
\end{figure}

\begin{figure}[h]
 \centering
 \includegraphics[width=0.7\textwidth]{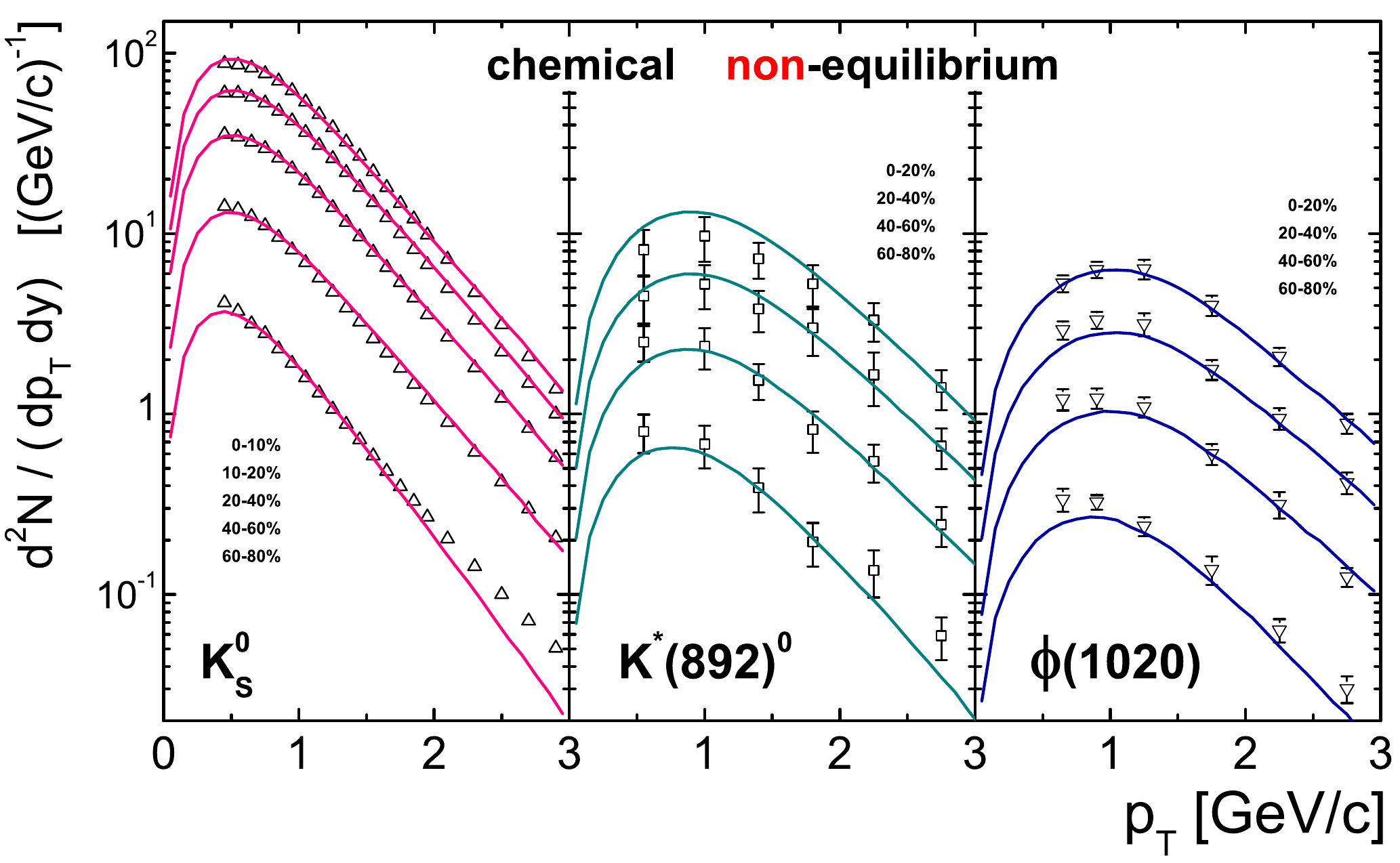}%K0S_Kstar_Phi}
 \caption{The spectra of strange particles in NEQ. The figure is from Ref.~\cite{Begun:2014rsa}.}
 \label{fig-2}
\end{figure}

There is only one additional parameter in the model that allows to
describe the spectra, because the product $\pi~\tau_f~r^2_{\rm
max}$ is equal to the volume per unit rapidity, while the ratio
$r_{\rm max}/\tau_f$ determines the slopes of the spectra.
The procedure is the following: first, the volume and temperature
are found from the fit of all mean multiplicities, and then the
$r_{\rm max}/\tau_f$ is found from the best fit to the spectra of
pions $\pi^++\pi^-$ and kaons $K^++K^-$. This approach gives
unexpectedly good results in NEQ. The fit that was made for the
spectra of pions and kaons only, appears also very good for
$p+\bar{p}$, $K^0_S$, $K^*(892)^0$ and $\phi(1020)$, see
Figs.~\ref{fig-1}, \ref{fig-2}.
The temperature in EQ is much higher, therefore the amount of
secondary pions is larger in NEQ. However, it is not enough to
reproduce the steepness of the spectrum at $p_T<250$~MeV, see
Fig.~\ref{fig-1} (right).
The best state of the art fit of pion, kaon and proton spectra in
EQ allows to describe protons taking into account rescattering
mechanism~\cite{Ryu:2015vwa}. Unfortunately, the authors of
Ref.~\cite{Ryu:2015vwa} do not provide the plot for pions in a
linear scale, or the data/model ratio. However, one can see that
they still have about 30\% deficit of pions in the model at
$p_T<250$~MeV.
The simultaneous fit of $K^*(892)^0$ and $\phi(1020)$ in NEQ is as
surprising, as the applicability of the pion-kaon fit to protons.
The $K^*(892)^0$ is short living, while $\phi(1020)$ is long
living. They should feel a long rescattering phase
differently~\cite{Knospe:2015nva}. The successful simultaneous
description of protons, pions, $K^*(892)^0$ and $\phi(1020)$ in
NEQ without rescattering may indicate that the rescattering phase
is well parameterized by the non-equilibrium parameters
(\ref{f-i}).

\begin{figure}[h]
 \centering
 \includegraphics[width=0.48\textwidth]{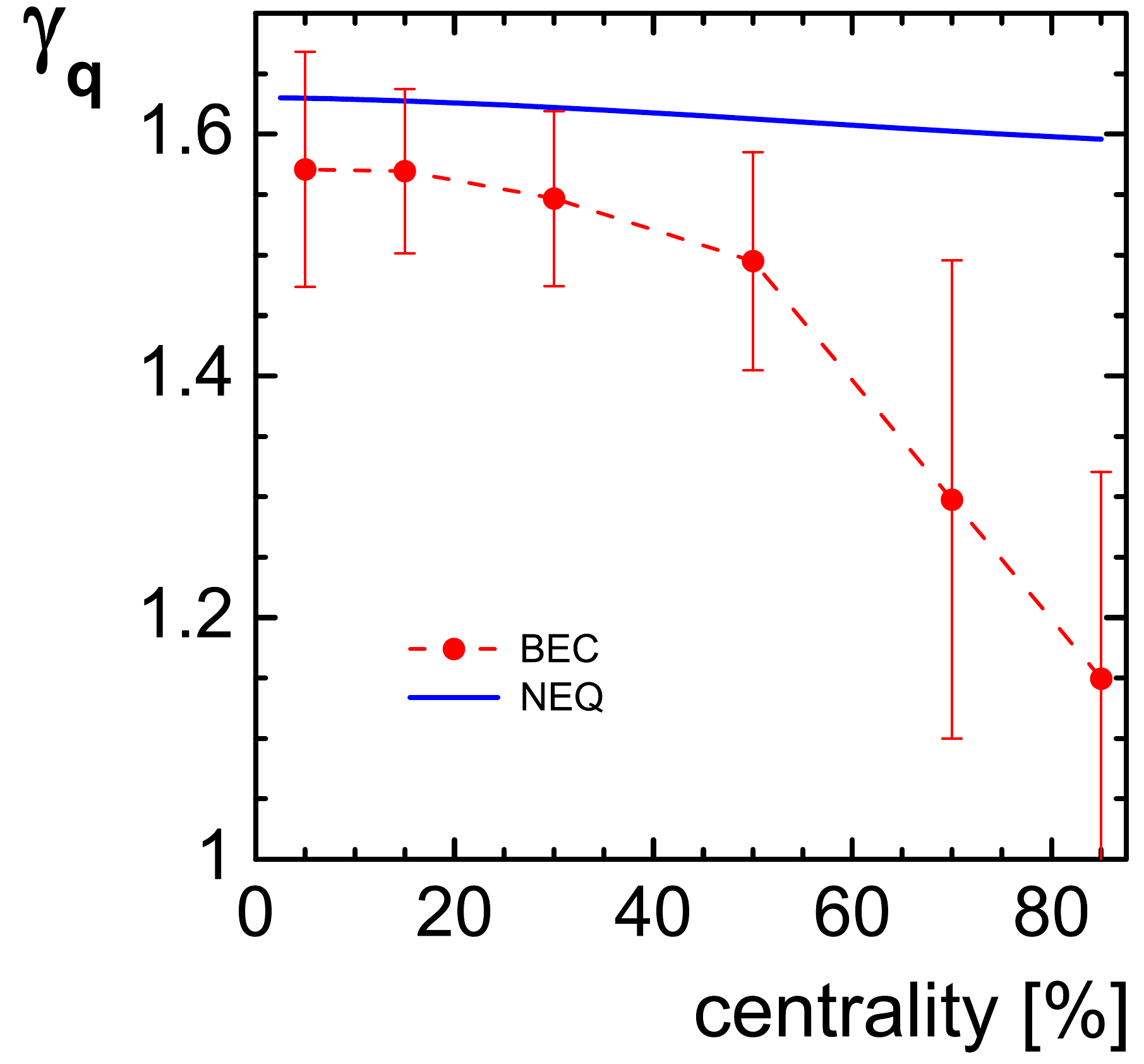}
 \hspace{0.2cm}
 \includegraphics[width=0.48\textwidth]{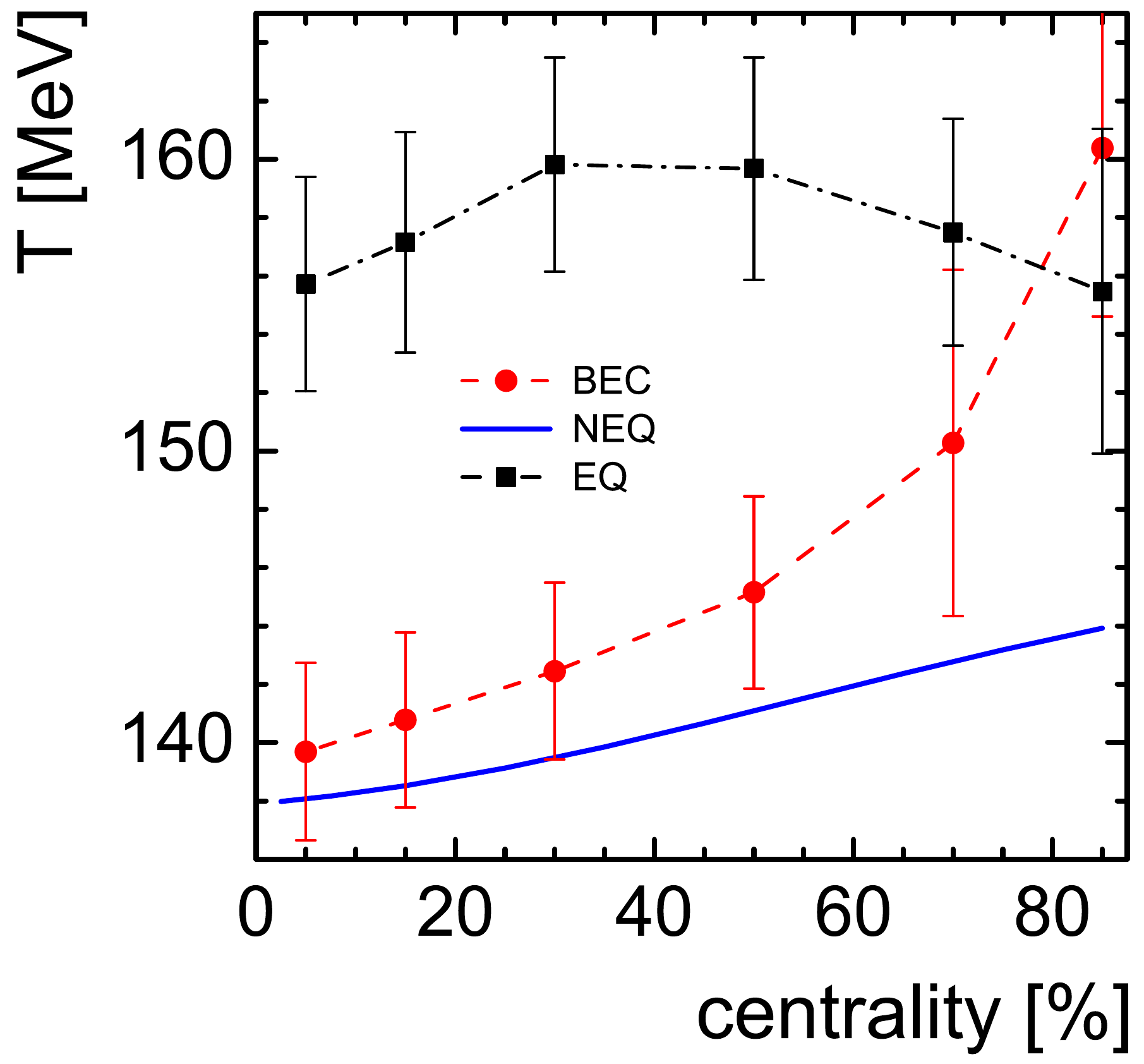}
 \caption{The non-equilibrium parameter $\gamma_q$ (left) and temperature (right) as the function of centrality.
 The figures are from Ref.~\cite{Begun:2014aha}}
 \label{fig-3}
\end{figure}

The numerical value of the $\gamma_q$ recalculated into the
corresponding chemical potential is very close to the pion mass:
$\mu_{\pi}=2T\ln\gamma_q\simeq 134~\text{MeV}~\simeq
m_{\pi^0}\simeq 134.98~$MeV. It suggests that a substantial part
of $\pi^0$ mesons tend to form the condensate, but this
possibility is not allowed in the HRG implementation in SHARE and
THERMINATOR. It can be done by explicit treatment of hadronic
ground states. When chemical potential approaches the mass of a
particle, $\mu\rightarrow m$, the discrete sum over the low
momentum quantum levels should be taken, instead of the integral
over momentum. It can be shown that in the infinite volume limit,
$V\rightarrow\infty$, only the zero momentum level, $p_0=0$, can
be taken into account~\cite{Begun:2008hq}. Then the total mean
multiplicity of a particle, $N_{\rm tot}$, receives one more term:
\eq{\label{Ncond}
 N_{\rm tot} &~=~ \sum_j \frac{1}{\exp[(\sqrt{p_j^2+m^2}-\mu)/T]-1}
 \nonumber \\
 &~\simeq~ \frac{1}{\exp[(m-\mu)/T]-1}
  +  V\int_0^{\infty} \frac{d^3p}{(2\pi)^3}\, \frac{1}{\exp[(\sqrt{p^2+m^2}-\mu)/T]-1}
  \\
 &~=~ N_{\rm cond} ~+~ N_{\rm norm}~,
 \nonumber
 }
where $N_{\rm cond}$ is the number of particles in the condensate.
When chemical potential approaches the mass, the $N_{\rm cond}$
grows infinitely and competes with arbitrarily large volume,
therefore the $\gamma_q$ does not grow so much. The corresponding
changes were introduced into SHARE. The new fit in the NEQ model
with explicit treatment of the ground state (BEC)\footnote{The
same abbreviation BEC is used for the Bose-Einstein condensation
and for the non-equilibrium model that allows the condensation on
the ground state.} gives rather different results, see
Fig.~\ref{fig-3}.
The inclusion of the ground state makes equilibrium and
non-equilibrium models closer. They even coincide at high
centrality, because $\gamma_q^{\rm BEC}\simeq 1$ there, while
$\gamma_q^{\rm EQ}=1$ by definition.

Pion spectra with the condensate were calculated in
Ref.~\cite{Begun:2015ifa} using the appropriately modified
THERMINATOR. The condensate is at rest only in it's reference
frame and is moving with the freeze-out hypersurface. The
approximation (\ref{Ncond}) leads to the step at $p_T\sim
230$~MeV, because the corresponding momentum distribution is given
by:
 \eq{
 \frac{dN}{dyd\phi_p\,p_Tdp_T}
 ~=~ \frac{N_{\rm cond}}{V}\,\frac{\tau_f^3}{m^2}~ \theta \left(r_{\rm max} - p_T\tau_f/m
 \right)~.
 }
where $\theta$ is the Heaviside step function.
The inclusion of several more low lying levels would lead to finer
steps that tend to a continuous line with increasing number of
steps, and have the same area under the curve. The current
approximation shows the maximal momentum that the condensate may
obtain from the movement of the freeze-out hypersurface, which is
$p_T<250$~MeV.

\begin{figure}[h]
 \centering
 \includegraphics[width=0.485\textwidth]{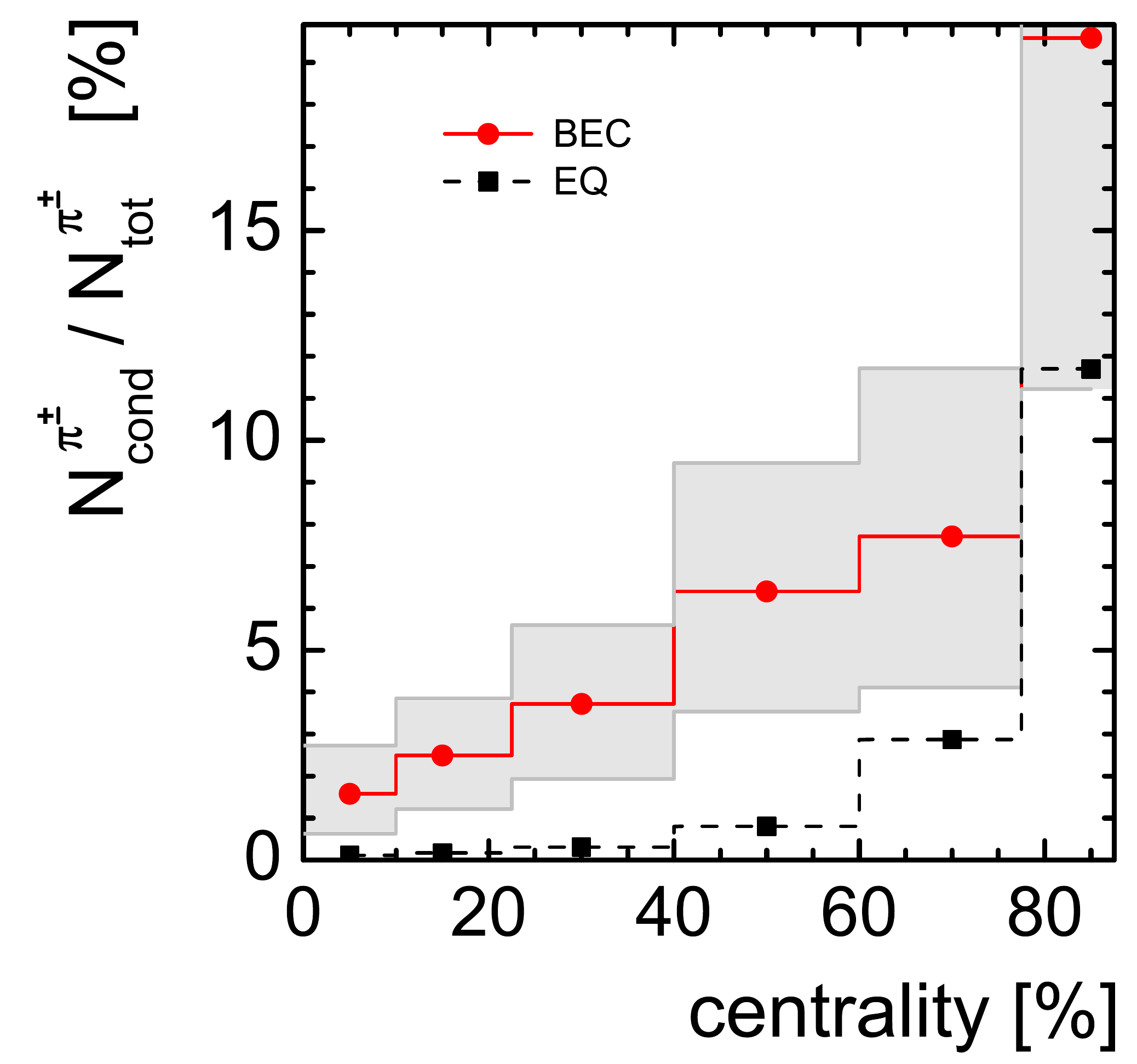}
 \hspace{0.2cm}
 \includegraphics[width=0.475\textwidth]{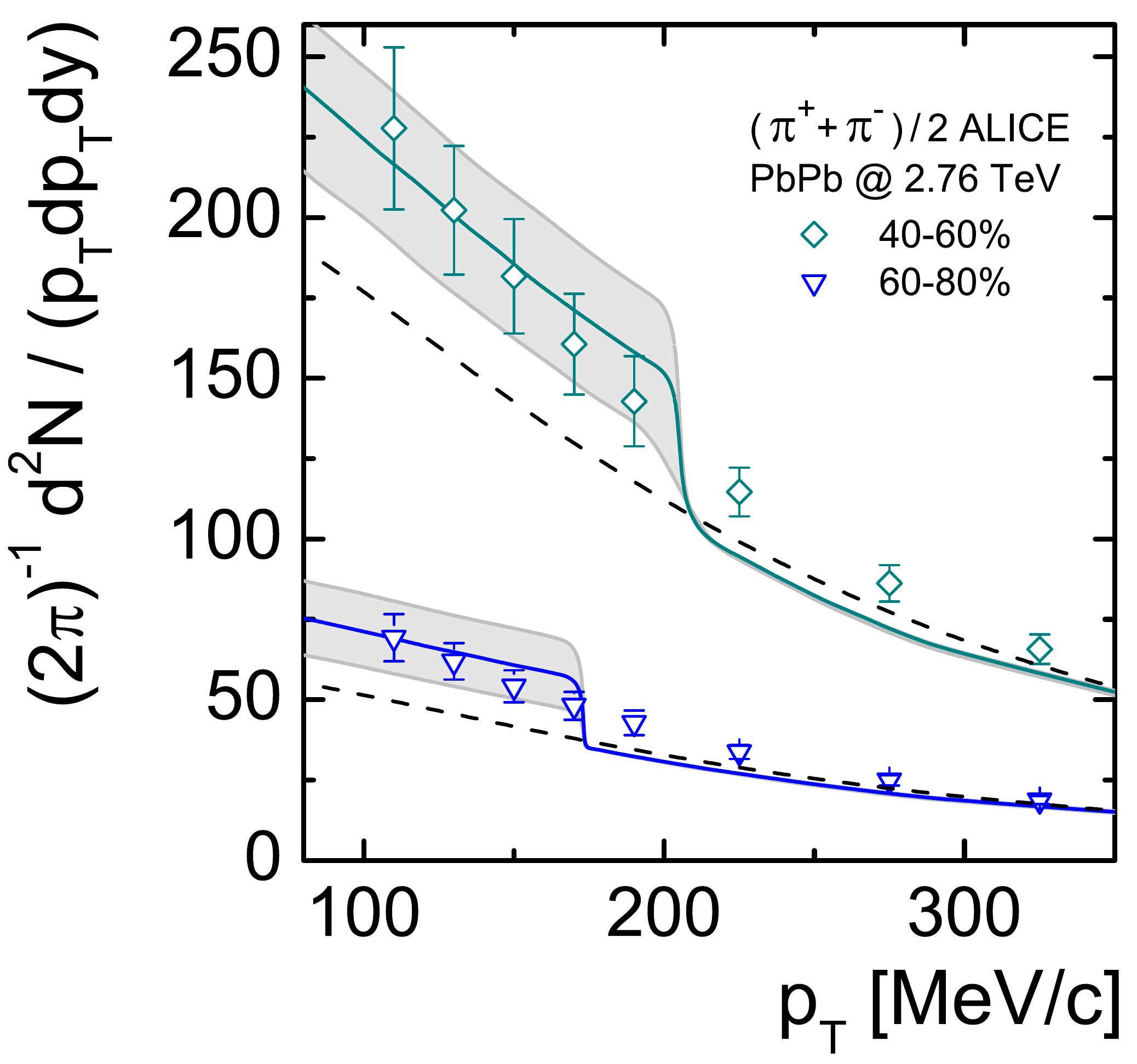}
 \caption{The pion spectra (left) and the condensate rate that was found in BEC (right) by the fit of
 mean multiplicities at the LHC. Solid line is for BEC, while dashed line is for EQ. The figures are from Ref.~\cite{Begun:2015ifa}.}
 \label{fig-4}
\end{figure}

The condensate rate grows with centrality in BEC, but the similar
growths shows EQ model, see Fig.~\ref{fig-4} (left). It is the
finite volume effect, which meas that the ground state
$N_0^{EQ}=N_{\rm cond}(\mu=0)$ should be taken into account even
in equilibrium for very peripheral, proton-proton, or
proton-nucleus collisions.
%
%The EQ fit for pions looks better than in Fig.~\ref{fig-1}, but one should remember that it still over-predicts protons at low $p_T$.
The gray area shows the 10\% deviation from the best fit. It is
well within the error bars at central, but not in peripheral
collisions, see Fig.~\ref{fig-4} (right). The combined data on
multiplicities and spectra are compatible with 5\% of pions in the
condensate~\cite{Begun:2015ifa}.

\begin{figure}[h]
 \centering
 \includegraphics[width=0.64\textwidth]{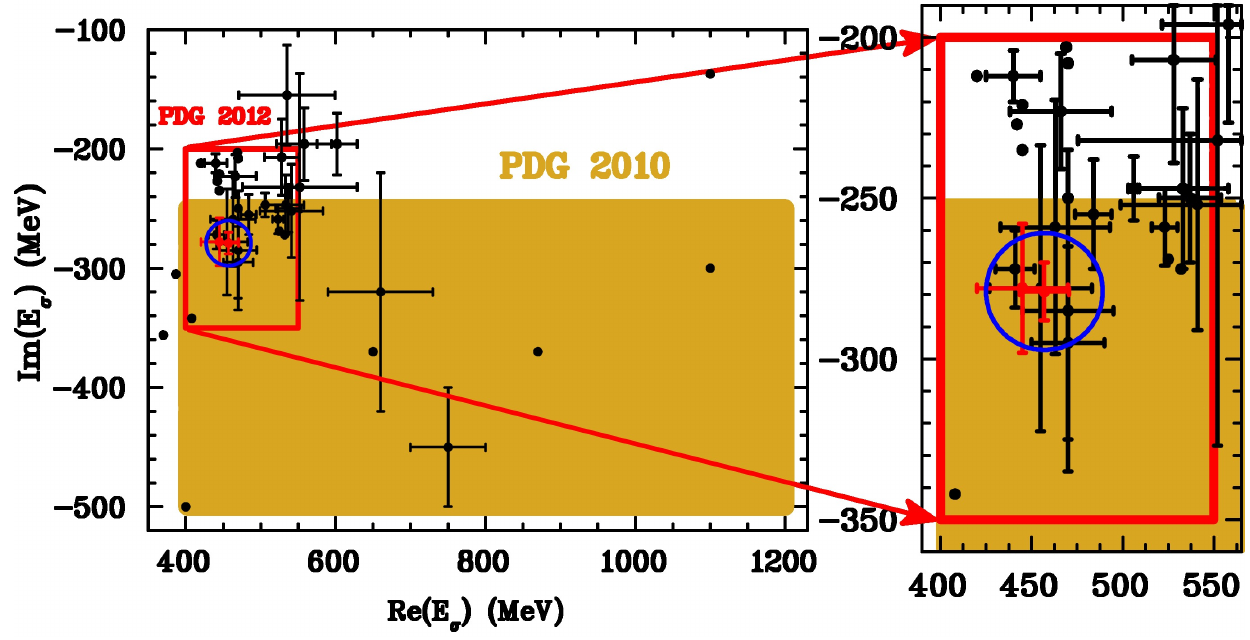}~~~
 %\hspace{0.2cm}
 \includegraphics[width=0.34\textwidth]{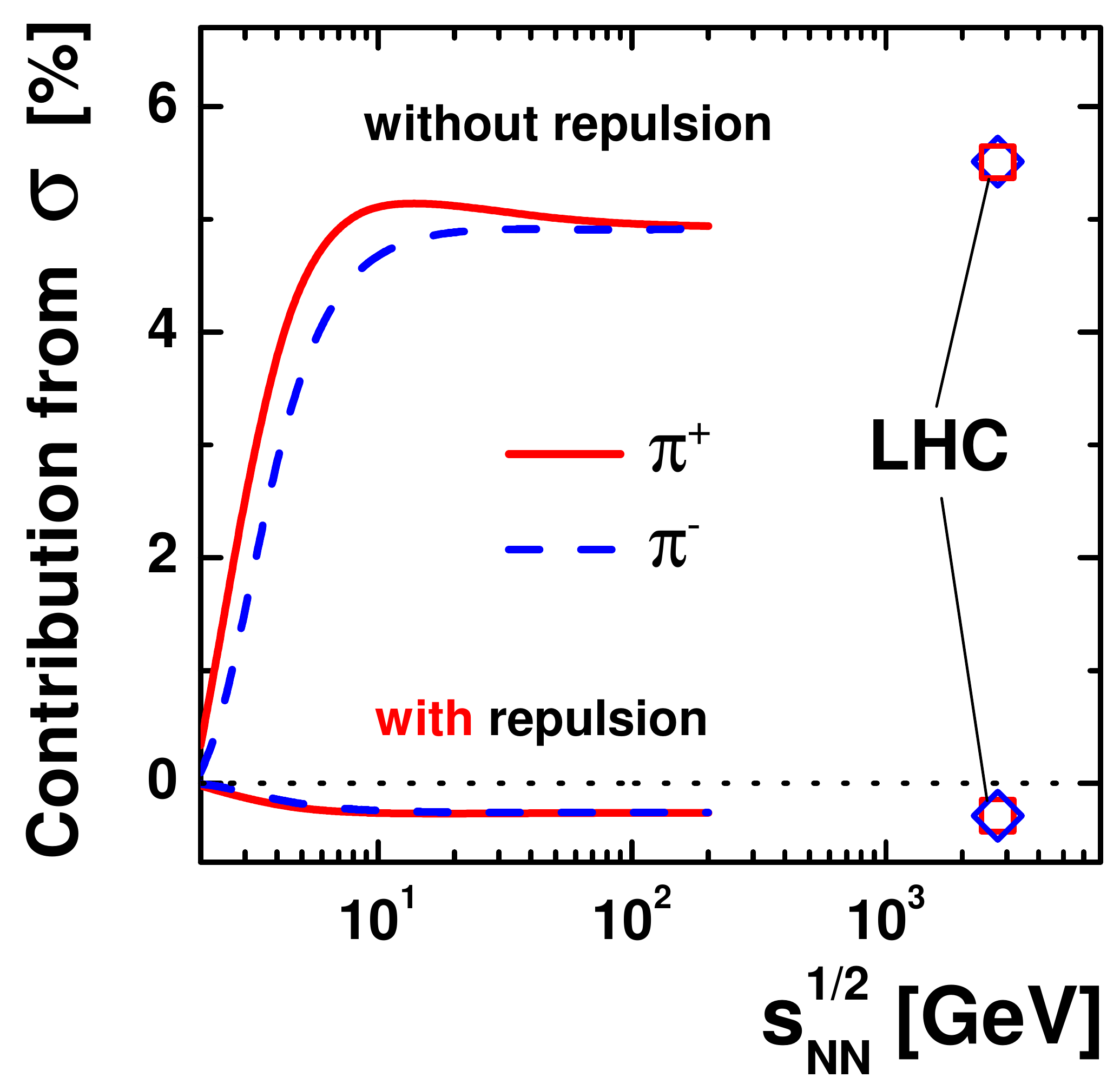}
 \caption{The change of the width and mass of the $\sigma$ meson since 2010 (left), and the contribution to pion yields from the decay of
 $\sigma$ (right). The left figure is from Ref.~\cite{Kaminski:2015sfa}.}
 \label{fig-5}
\end{figure}

The recent PDG reviews~\cite{Agashe:2014kda} report much lower
mass and width of the $f_0(500)$ resonance, or the sigma meson
($\sigma$), see Fig.~\ref{fig-5}, and
Ref.~\cite{GarciaMartin:2011jx,Pelaez:2015qba} for explanations.
Lower mass results in higher multiplicity in HRG. The $\sigma$
meson decays 100\% into pions, therefore the inclusion of the
updated $\sigma$ could have added some of the missing pions and
weaken the BEC signal.
The Breit-Wigner distribution is not accurate enough for so wide
resonances as $\sigma$. However, the width of the $\sigma$ can be
obtained from the derivative of the experimental $\pi\pi$ phase
shift, that we take from~\cite{GarciaMartin:2011jx}. It has
attractive (0;0) and repulsive (2;0) isospin-spin channel. The
attractive one is responsible for the emergence of the $f_0(500)$
resonance, however, the repulsive one cancels $f_0(500)$ until
$f_0(980)$ takes over above the mass $M_{\pi\pi}\sim 0.85~$GeV,
see Ref.~\cite{Broniowski:2015oha}.
%
%\begin{figure}[h]
% \centering
% \includegraphics[width=0.48\textwidth]{d0}
% \includegraphics[width=0.48\textwidth]{Rpi3}
% \caption{(Left) (Right) The figure is from Ref.~\cite{Broniowski:2015oha}.}
% \label{fig-6}
%\end{figure}
%
The cancellation happens on the level of the distribution
function, therefore it is present in all isospin-averaged
observables. The $\sigma$ implemented as a Breit-Winger pole with
$M_{\sigma}=484$ and $\Gamma_{\sigma}/2=255~$MeV produces up to
$5\%$ of pions, while the truth contribution is $-0.3\%$, see
Fig.~\ref{fig-5} (right)~\cite{Broniowski:2015oha}. The absence of
pions from the $\sigma$ decays enhances all ratios to pions,
compared to the results that were obtained in models with
$\sigma$. It is particularly important for the description of the
$K/\pi$ horn at the SPS~\cite{Gazdzicki:1998vd,Alt:2007aa} and the
proton-pion puzzle at the LHC.

\section{Conclusions}
The non-equilibrium thermal model combined with the single
freeze-out scenario explains very well the spectra of $\pi$, $K$,
$p$, $K^0_S$, $K^*(892)^0$ and $\phi(1020)$ particles at the LHC.
The introduction of the ground state decreases the non-equilibrium
parameters and increases the temperature with centrality.
The enhancement of the low $p_T$ pion spectra may be interpreted
as a signature of the onset of pion condensation at the LHC.
The missing pions from the $\sigma$ meson in HRG enhance the
proton-pion puzzle at the LHC, and allow for larger amount of the
condensate.
Many efforts and attention of the community is required to obtain
high temperature BEC and study it's properties. However, it could
open a wide new field of research.
\\
\\
{\bf Acknowledgments:}
\\
The author is thankful to W.~Florkowski,
M.~Rybczynski, M.~Gorenstein, W.~Broniowski and F.~Giacosa for
fruitful collaboration.
This work was supported by Polish National Science Center grant
No. DEC-2012/06/A/ST2/00390.

%
% BibTeX or Biber users please use (the style is already called in the class, ensure that the "woc.bst" style is in your local directory)
% \bibliography{name or your bibliography database}

\begin{thebibliography}{}
%
\bibitem{BEC}
 S.~N.~Bose, Z.\ Phys.\ {\bf 26}, 178 (1924);
 A.~Einstein, Sitz.\ Ber.\ Preuss.\ Akad.\ Wiss.\ (Berlin) {\bf 1}, 3 (1925).

\bibitem{BEC-1}
 M.~H.~Anderson, et al., Science {\bf 269}, 198 (1995); K.~B.~Davis, et al., Phys.\ Rev.\ Lett. {\bf 75}, 3969 (1995).

\bibitem{BEC-2}
 Editor Tore Frangsmyr, \textit{Les Prix Nobel}, (The Nobel Prizes 2001, Nobel Foundation, Stockholm,
 2002) 77-108.

\bibitem{Begun:2008hq}
  V.~V.~Begun and M.~I.~Gorenstein,
  %``Bose-Einstein Condensation in the Relativistic Pion Gas: Thermodynamic Limit and Finite Size Effects,''
  Phys.\ Rev.\ C {\bf 77}, 064903 (2008)
%  doi:10.1103/PhysRevC.77.064903
  [arXiv:0802.3349 [hep-ph]].

\bibitem{Abelev:2013vea}
  B.~Abelev {\it et al.} [ALICE Collaboration],
  %``Centrality dependence of $\pi$, K, p production in Pb-Pb collisions at $\sqrt{s_{NN}}$ = 2.76 TeV,''
  Phys.\ Rev.\ C {\bf 88}, 044910 (2013)
%  doi:10.1103/PhysRevC.88.044910
  [arXiv:1303.0737 [hep-ex]].

\bibitem{Stachel:2013zma}
  J.~Stachel, A.~Andronic, P.~Braun-Munzinger and K.~Redlich,
  %``Confronting LHC data with the statistical hadronization model,''
  J.\ Phys.\ Conf.\ Ser.\  {\bf 509}, 012019 (2014)
  doi:10.1088/1742-6596/509/1/012019
  [arXiv:1311.4662 [nucl-th]].

\bibitem{Cleymans:2014xha}
  J.~Cleymans,
  %``Status of the Thermal Model and Chemical Freeze-Out,''
  EPJ Web Conf.\  {\bf 95}, 03004 (2015)
%  doi:10.1051/epjconf/20159503004
  [arXiv:1412.7045 [hep-ph]].

\bibitem{Petran:2013lja}
  M.~Petran, J.~Letessier, V.~Petracek and J.~Rafelski,
  %``Hadron production and quark-gluon plasma hadronization in Pb-Pb collisions at $\sqrt{s_{NN}}=2.76$ TeV,''
  Phys.\ Rev.\ C {\bf 88}, no. 3, 034907 (2013)
%  doi:10.1103/PhysRevC.88.034907
  [arXiv:1303.2098 [hep-ph]].

\bibitem{Chatterjee:2014lfa}
  S.~Chatterjee, B.~Mohanty and R.~Singh,
  %``Freezeout hypersurface at energies available at the CERN Large Hadron Collider from particle spectra: Flavor and centrality dependence,''
  Phys.\ Rev.\ C {\bf 92}, no. 2, 024917 (2015)
%  doi:10.1103/PhysRevC.92.024917
  [arXiv:1411.1718 [nucl-th]].

\bibitem{Noronha-Hostler:2014aia}
  J.~Noronha-Hostler and C.~Greiner,
  %``Understanding the $p/\pi$ ratio at LHC due to QCD mass spectrum,''
  Nucl.\ Phys.\ A {\bf 931}, 1108 (2014)
%  doi:10.1016/j.nuclphysa.2014.08.101
  [arXiv:1408.0761 [nucl-th]].

\bibitem{Becattini:2012xb}
  F.~Becattini, M.~Bleicher, T.~Kollegger, T.~Schuster, J.~Steinheimer and R.~Stock,
  %``Hadron Formation in Relativistic Nuclear Collisions and the QCD Phase Diagram,''
  Phys.\ Rev.\ Lett.\  {\bf 111}, 082302 (2013)
%  doi:10.1103/PhysRevLett.111.082302
  [arXiv:1212.2431 [nucl-th]].

\bibitem{Shuryak:2014zxa}
  E.~Shuryak,
  %``Heavy Ion Collisions: Achievements and Challenges,''
  arXiv:1412.8393 [hep-ph].

\bibitem{Csorgo:1994dd}
  T.~Csorgo and L.~P.~Csernai,
  %``Quark - gluon plasma freezeout from a supercooled state?,''
  Phys.\ Lett.\ B {\bf 333}, 494 (1994)
%  doi:10.1016/0370-2693(94)90173-2
  [hep-ph/9406365].

\bibitem{Blaizot:2011xf}
  J.~P.~Blaizot, F.~Gelis, J.~F.~Liao, L.~McLerran and R.~Venugopalan,
  %``Bose--Einstein Condensation and Thermalization of the Quark Gluon Plasma,''
  Nucl.\ Phys.\ A {\bf 873}, 68 (2012)
%  doi:10.1016/j.nuclphysa.2011.10.005
  [arXiv:1107.5296 [hep-ph]].

\bibitem{Gelis:2014tda}
  F.~Gelis,
  %``Initial state in relativistic nuclear collisions and Color Glass Condensate,''
  Nucl.\ Phys.\ A {\bf 931}, 73 (2014)
%  doi:10.1016/j.nuclphysa.2014.09.028
  [arXiv:1412.0471 [hep-ph]].

\bibitem{Begun:2014rsa}
  V.~Begun, W.~Florkowski and M.~Rybczynski,
  %``Transverse-momentum spectra of strange particles produced in Pb+Pb collisions at $\sqrt{s_{\rm NN}}=2.76$ TeV in the chemical non-equilibrium model,''
  Phys.\ Rev.\ C {\bf 90}, no. 5, 054912 (2014)
%  doi:10.1103/PhysRevC.90.054912
  [arXiv:1405.7252 [hep-ph]].

\bibitem{Begun:2013nga}
  V.~Begun, W.~Florkowski and M.~Rybczynski,
  %``Explanation of hadron transverse-momentum spectra in heavy-ion collisions at $\sqrt s_{NN} =$ 2.76 TeV within chemical non-equilibrium statistical hadronization model,''
  Phys.\ Rev.\ C {\bf 90}, no. 1, 014906 (2014)
%  doi:10.1103/PhysRevC.90.014906
  [arXiv:1312.1487 [nucl-th]].

\bibitem{Begun:2015ifa}
  V.~Begun and W.~Florkowski,
  %``Bose-Einstein condensation of pions in heavy-ion collisions at the CERN Large Hadron Collider (LHC) energies,''
  Phys.\ Rev.\ C {\bf 91}, 054909 (2015)
  %  doi:10.1103/PhysRevC.91.054909
  [arXiv:1503.04040 [nucl-th]].

\bibitem{Abelev:2013pqa}
  B.~B.~Abelev {\it et al.} [ALICE Collaboration],
  %``Two- and three-pion quantum statistics correlations in Pb-Pb collisions at $\sqrt{{s}_{NN}} =$ 2.76 TeV at the CERN Large Hadron Collider,''
  Phys.\ Rev.\ C {\bf 89}, no. 2, 024911 (2014)
%  doi:10.1103/PhysRevC.89.024911
  [arXiv:1310.7808 [nucl-ex]].

\bibitem{Rafelski:2015cxa}
  J.~Rafelski,
  %``Melting Hadrons, Boiling Quarks,''
  Eur.\ Phys.\ J.\ A {\bf 51}, no. 9, 114 (2015)
%  doi:10.1007/978-3-319-17545-4_33, 10.1140/epja/i2015-15114-0
  [arXiv:1508.03260 [nucl-th]].

\bibitem{Torrieri:2004zz}
  G.~Torrieri, S.~Steinke, W.~Broniowski, W.~Florkowski, J.~Letessier and J.~Rafelski,
  %``SHARE: Statistical hadronization with resonances,''
  Comput.\ Phys.\ Commun.\  {\bf 167}, 229 (2005)
%  doi:10.1016/j.cpc.2005.01.004
  [nucl-th/0404083].

\bibitem{Chojnacki:2011hb}
  M.~Chojnacki, A.~Kisiel, W.~Florkowski and W.~Broniowski,
  %``THERMINATOR 2: THERMal heavy IoN generATOR 2,''
  Comput.\ Phys.\ Commun.\  {\bf 183}, 746 (2012)
%  doi:10.1016/j.cpc.2011.11.018
  [arXiv:1102.0273 [nucl-th]].

\bibitem{Broniowski:2001we}
  W.~Broniowski and W.~Florkowski,
  %``Explanation of the RHIC p(T) spectra in a thermal model with expansion,''
  Phys.\ Rev.\ Lett.\  {\bf 87}, 272302 (2001)
%  doi:10.1103/PhysRevLett.87.272302
  [nucl-th/0106050].

\bibitem{Begun:2014aha}
  V.~Begun,
  %``Thermal model for Pb+Pb collisions at $\sqrt{s}_{NN} = 2.76$ TeV with explicit treatment of hadronic ground states,''
  EPJ Web Conf.\  {\bf 97}, 00003 (2015)
%  doi:10.1051/epjconf/20159700003
  [arXiv:1412.6532 [nucl-th]].

\bibitem{Ryu:2015vwa}
  S.~Ryu, J.-F.~Paquet, C.~Shen, G.~S.~Denicol, B.~Schenke, S.~Jeon and C.~Gale,
  %``Importance of the Bulk Viscosity of QCD in Ultrarelativistic Heavy-Ion Collisions,''
  Phys.\ Rev.\ Lett.\  {\bf 115}, no. 13, 132301 (2015)
%  doi:10.1103/PhysRevLett.115.132301
  [arXiv:1502.01675 [nucl-th]].

\bibitem{Knospe:2015nva}
  A.~G.~Knospe, C.~Markert, K.~Werner, J.~Steinheimer and M.~Bleicher,
  %``Hadronic resonance production and interaction in partonic and hadronic matter in EPOS3 with and without the hadronic afterburner UrQMD,''
  arXiv:1509.07895 [nucl-th].

\bibitem{Agashe:2014kda}
  K.~A.~Olive {\it et al.} [Particle Data Group Collaboration],
  %``Review of Particle Physics,''
  Chin.\ Phys.\ C {\bf 38}, 090001 (2014).
%  doi:10.1088/1674-1137/38/9/090001

\bibitem{GarciaMartin:2011jx}
  R.~Garcia-Martin, R.~Kaminski, J.~R.~Pelaez and J.~Ruiz de Elvira,
  %``Precise determination of the f0(600) and f0(980) pole parameters from a dispersive data analysis,''
  Phys.\ Rev.\ Lett.\  {\bf 107}, 072001 (2011)
%  doi:10.1103/PhysRevLett.107.072001
  [arXiv:1107.1635 [hep-ph]].

\bibitem{Pelaez:2015qba}
  J.~R.~Pelaez,
  %``From controversy to precision on the sigma meson: a review on the status of the non-ordinary $f_0(500)$ resonance,''
  arXiv:1510.00653 [hep-ph].

\bibitem{Kaminski:2015sfa}
  R.~Kaminski,
  %``What happened with the $f_0(500)/\sigma $ meson? Theory and experiment,''
  Acta Phys.\ Polon.\ Supp.\  {\bf 8}, no. 1, 103 (2015).
%  doi:10.5506/APhysPolBSupp.8.103

\bibitem{Broniowski:2015oha}
  W.~Broniowski, F.~Giacosa and V.~Begun,
  %``Cancellation of the $\sigma$ meson in thermal models,''
  Phys.\ Rev.\ C {\bf 92}, no. 3, 034905 (2015)
%  doi:10.1103/PhysRevC.92.034905
  [arXiv:1506.01260 [nucl-th]].


\bibitem{Gazdzicki:1998vd}
  M.~Gazdzicki and M.~I.~Gorenstein,
  %``On the early stage of nucleus-nucleus collisions,''
  Acta Phys.\ Polon.\ B {\bf 30}, 2705 (1999)
  [hep-ph/9803462].

\bibitem{Alt:2007aa}
  C.~Alt {\it et al.} [NA49 Collaboration],
  %``Pion and kaon production in central Pb + Pb collisions at 20-A and 30-A-GeV: Evidence for the onset of deconfinement,''
  Phys.\ Rev.\ C {\bf 77}, 024903 (2008)
%  doi:10.1103/PhysRevC.77.024903
  [arXiv:0710.0118 [nucl-ex]].

\end{thebibliography}

\end{document}